\documentclass[12pt,a4paper]{article}

\newif\ifpdf
\ifx\pdfoutput\undefined
    \pdffalse                   
\else
    \ifx\pdfoutput\relax
        \pdffalse               
    \else
        \ifnum\pdfoutput>0
            \pdftrue            
        \else
            \pdffalse           
        \fi
    \fi
\fi

\ifpdf
    \usepackage[pdftex]{graphicx}
    \DeclareGraphicsExtensions{.pdf, .jpg, .tif}
\else
    \usepackage{graphicx}
    \DeclareGraphicsExtensions{.eps, .jpg}
\fi

\usepackage{amsfonts}
\usepackage{amssymb}
\usepackage{amsthm}
\usepackage{mathrsfs}
\usepackage{enumerate}
\usepackage{amsmath}
\usepackage{pgf}

\theoremstyle{plain}
\newtheorem{thm}{Theorem}
\newtheorem{prop}[thm]{Proposition}

\theoremstyle{customthm}
\newtheorem*{modass}{Model Assumptions}
\theoremstyle{definition}

\newcommand{\Sua}{S_{\scriptscriptstyle \textrm{U}|\textrm{A}}}
\newcommand{\Spua}{S_{\scriptscriptstyle \textrm{PU}|\textrm{A}}}
\newcommand{\taupua}{\tau_{\scriptscriptstyle \textrm{PU}|\textrm{A}}}
\newcommand{\Sa}{S_{\scriptscriptstyle \textrm{A}}}

\renewcommand{\P}{\mathbb{P}}
\newcommand{\Pnp}{\P_{{\scriptscriptstyle \textrm{NP}}}}
\newcommand{\Po}{P_{\textrm{o}}}
\newcommand{\No}{N_{\textrm{sp}}}
\newcommand{\Pt}{P_{\textrm{t}}}
\newcommand{\Ntot}{T_{\textrm{sp}}}
\newcommand{\dens}{f}
\newcommand{\densi}{f_{0}}

\newcommand{\E}[1]{\mathbb{E}\left[#1\right]}       

\newcommand{\figref}[1]{Figure~\ref{#1}}

\begin{document}

\title{On the Estimation of the Proportion of True Recent Infections Using the
BED Capture Enzyme Immunoassay}

\author{Thomas A. McWalter\footnote{School of Computational and Applied Mathematics, University of the Witwatersrand, South Africa} $ $ and Alex Welte\footnote{South African DST/NRF Centre of Excellence in Epidemiological Modelling and Analysis (SACEMA) and School of Computational and Applied Mathematics, University of the Witwatersrand, South Africa}}

\maketitle

\begin{abstract}
In this short note, we analyze the assumptions made by McDougal et al.~\cite{McDougaletal2006}, both explicit and implicit, in their estimation of the proportion of ``true recent infections'' using the BED CEIA. This enables us to write down expressions for the sensitivity, short term specificity and long term specificity of a test for recent infection defined by a BED ODn below a threshold. We then derive an identity which shows the relationship between these parameters, allowing the elimination of sensitivity and short term specificity from an expression relating the proportion of ``true recent infections'' to the proportion of seropositive individuals testing below threshold. This has two important consequences. Firstly, the simplified formula is substantially more amenable to calibration. Secondly, naively treating the parameters as independent would lead to an incorrect estimate of uncertainty due to imperfect calibration.
\end{abstract}

\newpage

\section*{Elimination of Parameters}

In the model proposed by McDougal et al.~\cite{McDougaletal2006}, a BED ODn below some threshold, for a seropositive individual, is declared to be an imperfect test for recent infection. They derive an estimate for the true proportion of recent infections ($\Pt$) in terms of the proportion of seropositive individuals that register under the threshold ($\Po$), a sensitivity ($\sigma$), a short term specificity ($\rho_1$) and a long term specificity ($\rho_2$). Knowledge of $\Pt$ allows the calculation of the `recent infections in 1 year per number at risk' in a hypothetical cohort. The present note concerns the correct calculation of $\Pt$, but does not address the issue of calculating a risk of infection using this proportion.

McDougal et al.~estimate a window period, being `the mean period of time from initial seroconversion to reaching an ODn of 0.8'. Presumably non-progressors---those not reaching the threshold---are censored. More specifically this implies that the window period is the mean threshold crossing time conditional on progression (i.e.~actually reaching the threshold). Sensitivity of the test is calculated for an interval corresponding to the window period. Short term specificity is calculated for `the interval immediately after, and equal in duration to, the window period'. Long term specificity is for `the period thereafter (where the curve is flat)'. The `curve' being referred to here is the survival function (for the calibration sample) in the state of being \emph{under} the threshold, conditional on being \emph{alive}, which we denote $\Sua(t)$. McDougal et al.~explicitly make the following assumptions, with the justification that they `are reasonable as very little attrition (from death) during the first two time intervals after infection would be expected':
\begin{enumerate}
    \item `Recent infections are randomly distributed within the first window period'.
    \item `The number of persons in the interval of equal duration immediately after the mean window period equals the number in the first window period'.
    \item `The remainder of the population is more than two window periods since seroconversion'.
\end{enumerate}

While it may be true in the situation being explored here, we note that it is not \emph{a priori} obvious that the choice of equal window periods ensures that $\Sua(t)$ is flat after twice the window period. With this in mind, we propose a generalization in which the two window periods be allowed to have arbitrary values $\omega_1$ and $\omega_2$, as long as all individuals that progress do so in a time less than $\omega_1+\omega_2$ after seroconversion (i.e.~$\Sua(t)$ is flat for $t>\omega_1+\omega_2$, see the bottom graph of Figure~1). For analytical convenience, we introduce $\Spua(t)$, the survival in the state of being under threshold for progressors (i.e.~those who reach the threshold). We also introduce $\Pnp$, the fraction of individuals that fail to progress. Then $\Sua(t)$, $\Spua(t)$ and $\Pnp$ are related by
\[
    \Sua(t)=(1-\Pnp)\Spua(t)+\Pnp.
\]
The introduction of $\Spua(t)$ allows us to provide a precise definition of the window period used by McDougal et al., being the mean period of time from initial seroconversion to reaching ODn spent by those who progress:
\[
    \omega:=\E{\taupua}=\int_0^\infty\Spua(t)\,dt,
\]
which follows from integration by parts on the relevant density function.

\input{SixSector.TpX}

Assumption~1 above means that infection times in the first window period are uniformly distributed. Although assumption~2 merely states that the \emph{number} of infections in the second window period is equal to the \emph{number} in the first, we shall see later that this is not strong enough to produce a definition of $\rho_1$ which is independent of the state of the sample. It is necessary to specify the stronger assumption that the infection events in the second window period are also \emph{uniformly distributed with the same intensity} as in the first window period. We see below that this assumption is implicit in the work of McDougal et al. To make this more tangible, we denote the density of infection times of individuals in the sample by $\dens(t)$, with the number of seropositive individuals given by $\No=\int_0^\infty\dens(t)\,dt$. Then in our model of general window periods $\dens(t)=\densi$ for all $t\in[0,\omega_1+\omega_2]$, and it follows that the ratio of infected people in the second window period to those in the first period is $\omega_2/\omega_1$. It should be noted that $\dens(t)$ depends on incidence, susceptible population and life expectancies over the history of the epidemic. With reference to \figref{sixsector}, we are now in a position to write expressions for the number of seropositive individuals in each sector:
\begin{align*}
    n_1&=\int_0^{\omega_1}\dens(t)(1-\Sua(t))\,dt\\
    &=\densi(1-\Pnp)\int_0^{\omega_1}(1-\Spua(t))\,dt\\
    n_2&=\int_0^{\omega_1}\dens(t)\Sua(t)\,dt\\
    &=\densi\omega_1\Pnp+\densi(1-\Pnp)\int_0^{\omega_1}\Spua(t)\,dt\\
    n_3&=\int_{\omega_1}^{\omega_1+\omega_2}\dens(t)(1-\Sua(t))\,dt\\
    &=\densi(1-\Pnp)\int_{\omega_1}^{\omega_1+\omega_2}(1-\Spua(t))\,dt\\
    n_4&=\int_{\omega_1}^{\omega_1+\omega_2}\dens(t)\Sua(t)\,dt\\
    &=\densi\omega_2\Pnp+\densi(1-\Pnp)\int_{\omega_1}^{\omega_1+\omega_2}\Spua(t)\,dt\\
    n_5&=\int_{\omega_2}^\infty\dens(t)(1-\Sua(t))\,dt\\
    &=(1-\Pnp)\int_{\omega_2}^\infty\dens(t)\,dt\\
    n_6&=\int_{\omega_2}^\infty\dens(t)\Sua(t)\,dt\\
    &=\Pnp\int_{\omega_2}^\infty\dens(t)\,dt.
\end{align*}
Using the above expressions, the sensitivity, the short-term specificity and the long-term specificity are given by
\begin{align*}
    \sigma&=\frac{n_2}{n_1+n_2}=\frac{(1-\Pnp)\int_0^{\omega_1}\Spua(t)\,dt+\omega_1\Pnp}{\omega_1}\\
    \rho_1&=\frac{n_3}{n_3+n_4}=\frac{(1-\Pnp)\int_{\omega_1}^{\omega_1+\omega_2}(1-\Spua(t))\,dt}{\omega_2}\\
    \rho_2&=\frac{n_5}{n_5+n_6}=1-\Pnp.
\end{align*}

We can now see why the assumption of uniform distribution of infection events for the first and second window periods is required---it is the only way in which we can get a cancelation of $\dens(t)$ in the expressions for $\sigma$ and $\rho_1$.

We also see why it is necessary that $\Sua(t)$ must be flat after both window periods---it ensures that the $\Sua(t)$ is constant and can be pulled out of the integrals in $n_5$ and $n_6$ as the factor $\Pnp$. This is necessary for $\rho_2$ to be independent of $\dens(t)$.

Furthermore, we now show that in order to specify $\rho_2$ so that it is independent of the state of the epidemic, an implicit assumption is being made that survival is the same for progressors and non-progressors. Under bias-free recruitment into the survey, we have
\[
    \dens(t)=\frac{\No}{\Ntot}H(-t)I(-t)\Sa(t),
\]
where $H(t)$ is the number of healthy (susceptible) individuals, $I(t)$ is the instantaneous incidence, $\Sa(t)$ is the life-expectancy survival function measured from the time since infection and
\[
    \Ntot=\int_0^\infty H(-t)I(-t)\Sa(t)\,dt
\]
is the total number of seropositive individuals alive in the population at $t=0$. The ratio $\No/\Ntot$ is just the fraction of the total population that has been recruited into the survey. Now, note that $\dens(t)$ is used symmetrically in the expressions for $n_5$ and $n_6$. If different life expectancies were used in these formulae, reflecting a difference in survival for progressors and non-progressors, the $\dens$s in these formulae would need to be different, and would not cancel in the expression for $\rho_2$. This assumption is not explicitly stated by McDougal et al.~but is implicit in arriving at a $\rho_2$ that is independent of epidemic state.

With the calibration parameters specified precisely, we now derive an estimate for the proportion of seropositive people $\Pt$ who were infected at a time less than $\omega_1$ before the present---these are the \emph{true} recent infections. A generalization of equation (1) in McDougal et al., relating the proportion of true recent infections to the observed proportion of the population that tested recent, is given by
\[
    \Po=\Pt\sigma+\Pt\frac{\omega_2}{\omega_1}(1-\rho_1)+\left(1-\Pt-\Pt\frac{\omega_2}{\omega_1}\right)(1-\rho_2).
\]
This means that we can solve for the true proportion of recently infected individuals to get
\begin{equation}\label{Ptlabel}
    \Pt=\frac{\Po+\rho_1-1}{\sigma-\frac{\omega_2}{\omega_1}\rho_1+\left(1+\frac{\omega_2}{\omega_1}\right)\rho_2-1}.
\end{equation}
Note that this equation reduces to the one derived by McDougal et al.~when one sets $\omega_1=\omega_2$
\begin{equation}\label{badform}
    P_t=\frac{\Po+\rho_2-1}{\sigma-\rho_1+2\rho_2-1}.
\end{equation}

Now, for completeness, we provide the precise assumptions that are required in order to facilitate the analysis in the rest of this paper. We note that with the exception of arbitrary sized window periods, these assumptions are equivalent to the assumptions---either explicit or implicit---that are being made by McDougal et al.

\begin{modass}
Specify window periods $\omega_1$ and $\omega_2$. We assume that:
\begin{enumerate}
    \item The window periods are chosen so that the survival function \emph{$\Sua(t)$} is flat after $t=\omega_1+\omega_2$. This means that \emph{$\Spua(t)$} only has support on the time interval $t\in[0,\omega_1+\omega_2]$.
    \item Arrival times of infection events are uniformly distributed on the interval $[0,\omega_1+\omega_2]$.
    \item Survival is symmetric for progressing and non-progressing individuals.
\end{enumerate}
\end{modass}

We are now able to provide an important identity that is not anticipated in McDougal et al.
\begin{prop}
Under the model assumptions stated above, the following identity holds:
\[
    \sigma-\frac{\omega_2}{\omega_1}\rho_1+\left(1+\frac{\omega_2}{\omega_1}-
    \frac{\omega}{\omega_1}\right)\rho_2=1.
\]
\end{prop}

\begin{proof}
Since we assume that $\Spua(t)$ only has support on $t\in[0,\omega_1+\omega_2]$, we have
\[
    \int_0^{\omega_1+\omega_2}\Spua(t)\,dt=\int_0^\infty\Spua(t)\,dt=\E{\taupua}=\omega.
\]
Then, simply evaluating
\begin{align*}
    \sigma-\frac{\omega_2}{\omega_1}\rho_1&=\frac{(1-\Pnp)\int_0^{\omega_1}\Spua(t)\,dt+\omega_1\Pnp}{\omega_1}\\
    &\qquad-\frac{\omega_2}{\omega_1}\frac{(1-\Pnp)\int_{\omega_1}^{\omega_1+\omega_2}(1-\Spua(t))\,dt}{\omega_2}\\
    &=\frac{(1-\Pnp)\int_0^{\omega_1+\omega_2}\Spua(t)\,dt-\int_{\omega_1}^{\omega_1+\omega_2}(1-\Pnp)\,dt+\omega_1\Pnp}{\omega_1}\\
    &=\frac{(1-\Pnp)(\omega-\omega_2-\omega_1)+\omega_1}{\omega_1}\\
    &=1-\left(1+\frac{\omega_2}{\omega_1}-\frac{\omega}{\omega_1}\right)\rho_2,
\end{align*}
yields the result directly.
\end{proof}

Using the Proposition, equation \eqref{Ptlabel} simplifies to
\[
    \Pt=\frac{\omega}{\omega_1}\frac{\Po+\rho_2-1}{\rho_2}.
\]
This expression no longer relies on estimates for $\sigma$ and $\rho_1$. It is also interesting to note that it does not depend on $\omega_2$. Furthermore, if we set $\omega_1=\omega$ as in McDougal et al.~then we get
\begin{equation}\label{goodform}
    P_t=\frac{\Po+\rho_2-1}{\rho_2}.
\end{equation}

\section*{Discussion}

Note that \eqref{badform} as stated in McDougal et al.~contains three calibration parameters ($\sigma$, $\rho_1$ and $\rho_2$), while \eqref{goodform} contains only one calibration parameter ($\rho_2$). Incidence estimates using \eqref{goodform} would still, however, require the estimation of $\omega$. The method of McDougal et al.~can in principle be applied to an arbitrarily declared window period, as long as $\sigma$, $\rho_1$ and $\rho_2$ are calibrated for that value. We have therefore reduced the number of calibration parameters by one.

Estimation of extra parameters may unnecessarily dilute the statistical power of the calibration data at hand. Moreover, estimates of the uncertainty due to calibration, based on the assumption of the independence of $\sigma$, $\rho_1$ and $\rho_2$, will produce incorrect results. Note that when one sets  $\omega_1=\omega_2=\omega$, that the identity is given by
\[
    \sigma-\rho_1+\rho_2=1.
\]
Substituting the values for the calibrated parameters found by McDougal et al., namely $\sigma=0.768$, $\rho_1=0.723$ and $\rho_2=0.944$, into this equation gives a value for the left hand side of the equation equal to 0.989. This confirms that their calibration was reasonably accurate.

Perhaps the most important advantage of eliminating $\sigma$ and $\rho_1$ is that the remaining parameters are more amenable to calibration. The calibration of $\sigma$ and $\rho_1$ requires obtaining specimens from individuals with confidence about their time since infection (i.e.~using frequent follow-up). On the other hand both $\rho_2$ and $\omega$ can be estimated through follow-up intervals greater than $\omega_1+\omega_2$. The estimate for $\rho_2$ comes from the proportion of under-threshold samples known to be obtained more than $\omega_1+\omega_2$ since infection (i.e.~second seropositive samples). Given an estimate for $\rho_2$, $\omega$ can be estimated from the fraction of individuals who test under-threshold on the first seropositive sample.

We have suggested an alternative incidence estimation paradigm \cite{McWalterWelte2008} which requires fewer assumptions than the method of McDougal et al. In this approach $\Pnp=1-\rho_2$ and $\omega$ emerge as the natural calibration parameters.

\bibliography{parameters}
\bibliographystyle{plain}
\end{document}